\documentclass[12pt, twoside, here]{article}
\usepackage{epsf}
\usepackage{times,colordvi,amsmath,epsfig,float,color,multicol}
\usepackage{graphics}
\usepackage{hhline}
\usepackage[large]{subfigure}
\usepackage[latin1]{inputenc}
\usepackage{listings}
\usepackage{rotating}

\title{Could Deficiencies in South African Data Be the Explanation for Its Early \mbox{SARS-CoV-2} Peak?}

\author{S. J. Childs \\ \\ {\small\em Department of Mathematics and Applied Mathematics , University of Fort Hare,} \\ {\small\em Private Bag X1314, Alice, 5700, South Africa}}

\date{18 August, 2020}       
\begin{document}

\maketitle

\begin{abstract}
\noindent {\em The \mbox{SARS-CoV-2} pandemic peaked very early in comparison to
the thresholds predicted by an analysis of prior lockdown regimes. The most
convenient explanation is that some, external factor changed the value of the
basic reproduction number, $r_0$; and there certainly are arguments for this.
Other factors could, nonetheless, have played a role. This research attempts to
reconcile the observed peak with the thresholds predicted by lockdown regimes
similar to the one in force at the time. It contemplates the effect of two,
different, hypothetical errors in the data: The first is that the true level of
infection has been underestimated by a multiplicative factor, while the second
is that of an imperceptible, pre-existing, immune fraction of the population.
While it is shown that it certainly is possible to manufacture the perception of
an early peak as extreme as the one observed, solely by way of these two
phenomena, the values need to be fairly high. The phenomena would not, by any
measure, be insignificant. It also remains an inescapable fact that the early
peak in infections coincided with a fairly profound change in $r_0$; in all the
contemplated scenarios of data-deficiency.} 
\end{abstract}


Declaration of interest: None.

\section{Introduction}

%$0.7\%$ $0.6\%$ 20 July  $0.8\%$  26 July
On around the 18th of July, 2020, the incidence of \mbox{SARS-CoV-2} cases
peaked and the number of active infections followed suit around five days later.
Either some unknown factor caused the basic reproduction number, $r_0$, to drop
below unity, around the 13th of July (the mean \mbox{SARS-CoV-2} incubation
period is 5.2 days \cite{meanIncubationPeriod}), or the threshold had been
reached. All this transpired at an apparently very low level of total infection,
a mere 0.7 \% of the population; or, couching this in the more conventional
terms of susceptability, 99.3 \%. Such a threshold would imply a basic
reproductive number less than 1.01. On the 13th of July, South Africa did revert
to a lockdown regime similar to its previous level 4 lockdown ($r_0 =$ 1.69),
referred to in this work as ``level 3.5'', however, as much as prohibition,
curfews and a number of other measures were reinstated, the threshold predicted
by an analysis of the level 4 lockdown regime, suggests that a new regime, in
itself, was not enough to cause the observed peak. A basic reproductive number
of 1.01 is exceptionally marginal. How does one explain this conundrum?  

It is already known that the perception of a peak at 99.3 \% is based on
infection-data which are deficient by an order of magnitude, or even greater. 
The head of the CDC, Robert Redfield's opinion on the topic of asymptomatic or
undiagnosed \mbox{SARS-CoV-2} infections, in the U.S.A., is that antibody
testing reveals that ``A good rough estimate now is 10 to 1'' (\cite{redfield})
and others, in similar positions all over the world, have expressed similar
sentiments. Redfield's factor of eleven also needs to be revised upward if one
considers that, although antibodies lend themselves favourably to the diagnosis
of immunity, they are not the ultimate indicator. Undetected, T-cell mediated
immunity can exist in the absence of a positive antibody test. In South Africa,
epidemiologists have focussed on excess deaths and put forward a value of 1.59
(\cite{Dorrington}, \cite{excessDeaths1} and \cite{excessDeaths2}). There is not
necessarily any conflict between this apparently, relatively low number of
excess deaths and Redfield's statement, if one considers the obvious bias in
detection: If you're so sick that you're about to die, you're more likely to
seek out medical assistance and be diagnosed. It may also be worth keeping in
mind that a massive 57 \% of the inhabitants of the Mumbai slum areas of
Chembur, Matunga and Dahisar tested positive for exposure to \mbox{SARS-CoV-2}
\cite{mumbai}. That the \mbox{SARS-CoV-2} virus is often insidious and infection
data are consequently incorrect by a factor is therefore already a widely
recognised phenomenon. This fact, alone, is nonetheless unable to reasonably
explain the \mbox{SARS-CoV-2} threshold observed in the South African data,
without contemplating improbably-high, though not impossible, values. 

The question then arises as to whether the \mbox{SARS-CoV-2} virus really is
novel and the population really is naive, or whether some other pathogen,
genetics, etc. has not imparted an undetectable immunity to a significant
fraction of the population. Although antibodies lend themselves favourably to
the diagnosis of immunity, they are not the ultimate indicator. As already
stated, undetected, T-cell mediated immunity can exist in the absence of a
positive antibody test. 

This research attempts to reconcile the observed peak in South African
infections with the thresholds predicted by lockdown regimes similar to the one
in force at the time. It contemplates the effect of two, different, hypothetical
errors in the data: The first is that the true level of infection has been
underestimated by a multiplicative factor, denoted $a$, while the second is that
of an imperceptible, pre-existing, immune fraction of the population, denoted
$b$. Since the rules in place at the time of the peak were most similar to level
4, the values of $a$ and $b$ explored were mostly selected on the basis that
they manufacture an erroneously-detected, \mbox{99.3 \%} threshold for the level
4 lockdown. 

Of course, it is possible that some, other, external factor, caused the basic reproduction number to plummet, around the 13th of July, and a quantification of lockdown regimes means nothing if the public at large were not compliant with the rules.

\section{The Erroneous Perception Created by Inexplicable Immunity and Asymptomatic, or Undiagnosed, Infections}

Suppose that $S(t)$, $I(t)$ and $R(t)$ represent the usual quantities in Kermack
and McKendrick's SIR model \cite{kermackAndMcKendrick}. Suppose that a tilde is
used to further distinguish detected values of these quantities, which for the
purposes of this exposition, will be erroneous. If, for some presently
inexplicable reason, there is an imperceptible, pre-existing immune fraction of
the population, $b$, then $R(0) = {\tilde R}(0) + b$, for an epidemic that
begins at some time, \mbox{$t=0$}. Suppose one were to further determine that
both ${\tilde I}(t)$ and therefore, the resistant portion arising from the
current epidemic, ${\tilde R}(t) - {\tilde R}(0)$, are in actual fact higher by
some factor, $a$. That is,  
\[
I(t) = a {\tilde I}(t) \hspace{10mm} \mbox{and} \hspace{10mm} R(t) = a \left[ {\tilde R}(t) - {\tilde R}(0) \right] + {\tilde R}(0) + b. 
\]
Substituting these quantities into
\begin{eqnarray} \label{realS} 
S(t) &=& 1 - I(t) - R(t) \nonumber \\
&=& 1 - a \left[ {\tilde I}(t) + \left( {\tilde R}(t) - {\tilde R}(0) \right)  \right] - {\tilde R}(0) - b \nonumber \\
&=& 1 - a + a \left[ 1 - \left( {\tilde I}(t) + {\tilde R}(t) \right) \right] - ( 1 - a ) {\tilde R}(0) - b \nonumber \\
&=& a {\tilde S}(t) + ( 1 - a ) [ 1 - {\tilde R}(0) ] - b, 
\end{eqnarray}
in which \mbox{${\tilde S}(t) = 1 - {\tilde I}(t) - {\tilde R}(t)$}. 

In the particular case of the South African, \mbox{SARS-CoV-2} peak, ${\tilde S}(t) = 0.993$ and ${\tilde R}(0) = 0$. Equation (\ref{realS}) therefore simplifies to
\begin{eqnarray} \label{realS2}
S(t) &=& 1 - 0.007 \times a - b.
\end{eqnarray}
The true value of the erroneously-detected 99.3 \% threshold therefore depends on the values of both $a$ and $b$, some examples of which are given in Table \ref{thresholds}.
\begin{table}
\begin{center}
\begin{tabular}{|c|c|c|c|} \hline & & & \\ Detected Threshold, ${\tilde S} \ / \%$ & \ $a$ \ & \ $b$ \ & True Value, $S \ / \%$ \\ & & & \\ \hline 
& & & \\ 
99.3 & 1.0 & 0.1 & 89.3 \\ 
& & & \\ 
\hline 
& & & \\ 
\texttt{"} & 10 & 0.0 & 93.0 \\ 
& & & \\ 
\hline 
& & & \\ 
\texttt{"} & 10 & 0.83 & 10.0 \\ 
& & & \\ 
\hline 
& & & \\ 
\texttt{"} & 20 & 0.65 & 21.0 \\ 
& & & \\ 
\hline 
& & & \\ 
\texttt{"} & 30 & 0.5 & 21.0 \\ 
& & & \\ 
\hline 
& & & \\ 
\texttt{"} & 60 & 0.0 & 58.0 \\ 
& & & \\ 
\hline 
\end{tabular}
\caption{Some examples of what the true value of an erroneously-detected peak of
99.3 \% would be, based on the substitution of various values of $a$ and $b$
into Equation (\ref{realS2}).}
\label{thresholds}
\end{center}
\end{table}

Changing the subject of Equation (\ref{realS}),
\begin{eqnarray} \label{perceivedS} 
{\tilde S}(t) &=& \frac{1}{a} \left[ S(t) + ( a - 1 ) [ 1 - {\tilde R}(0) ] + b \right]. 
\end{eqnarray}
This is a formula for the perceived susceptable fraction that will be erroneously detected for given values of $a$ and $b$.

\section{The Data and Their Interpretation}

Epidemiological data are usually presented in the format ``numbers of current
infections'' and ``total number of cases''. The present case of the
\mbox{SARS-CoV-2} pandemic is no exception. The data for the level-5 lockdown,
the level-4 lockdown, the early level-3 lockdown and Sweden were already
interpreted in \cite{childs} and may be found in Table \ref{results}. The final
level-3 and so-called level-3.5 values, in Table \ref{results}, were determined
as follows. 

\subsection*{The Level-3 Lockdown}

\begin{figure}[H]
    \begin{center}
        \includegraphics[width=15cm, angle=0, clip = true]{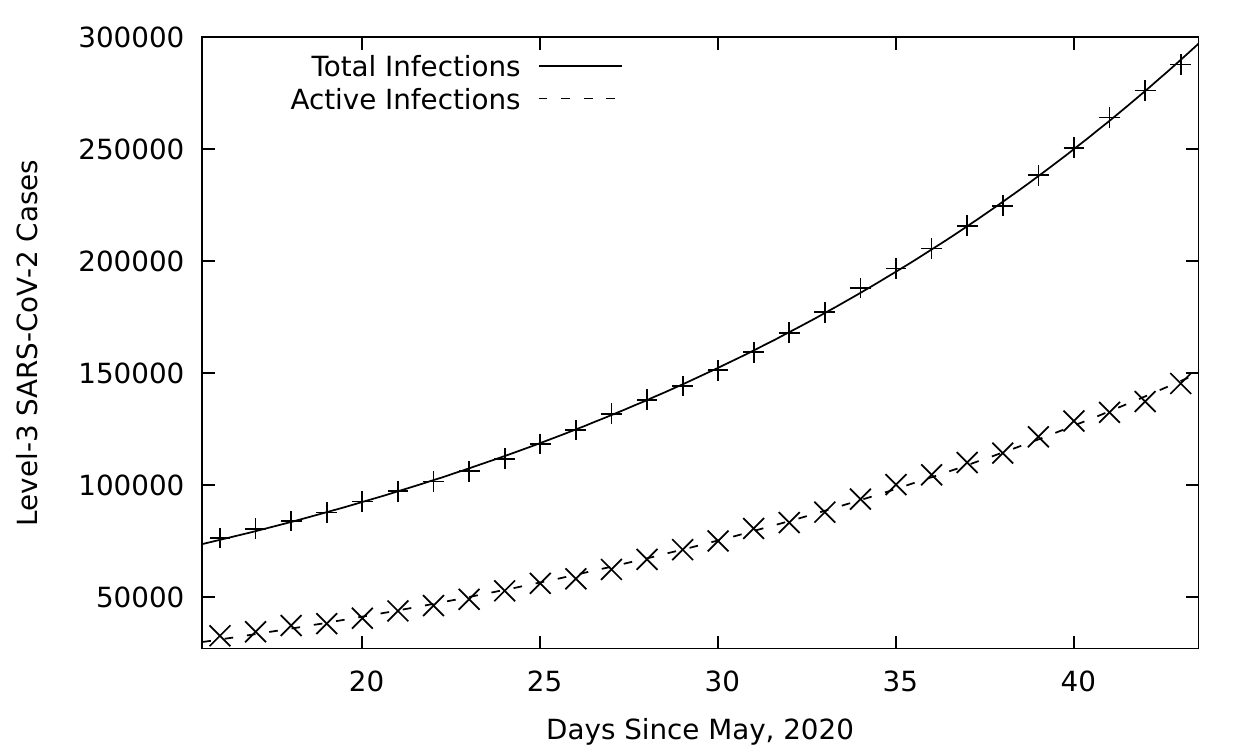}
	\caption{Level-3, best fits to \mbox{SARS-CoV-2}, infection data (16th of June to the 13th of July, 2020).} \label{level3}
   \end{center}
\end{figure}

The level-3 lockdown commenced on the 1st of June and was still in force up until the 12th of July. Once again, a period of 15 days was allowed for the viral incubation period and the subsequent diagnosis of an infection. Once again, it was also assumed that the termination of the level-3 lockdown would not reflect in the data for at least 24 hours. Curves were accordingly fitted to the subset of data (\cite{data}, \cite{openingWeeks} and \cite{data3}) which commenced on the 16 of June and terminated on the 13th of July. The curves fitted to the data, using Gnuplot, are depicted in Figure \ref{level3}.
The formula for the ``total infections'' curve is 
\[ 
35963.9 \ e^{0.048776 \ t} - \ 2993.68
\] 
and the formula for the ``active infections'' curve is 
\[ 
29735 \ e^{0.040912 \ t} - \ 26225.1. 
\] 
The values these formulae yield for the relevant dates are provided in \mbox{Table \ref{results}}. 

\subsection*{The Level-3.5 Lockdown}

\begin{figure}[H]
    \begin{center}
        \includegraphics[width=15cm, angle=0, clip = true]{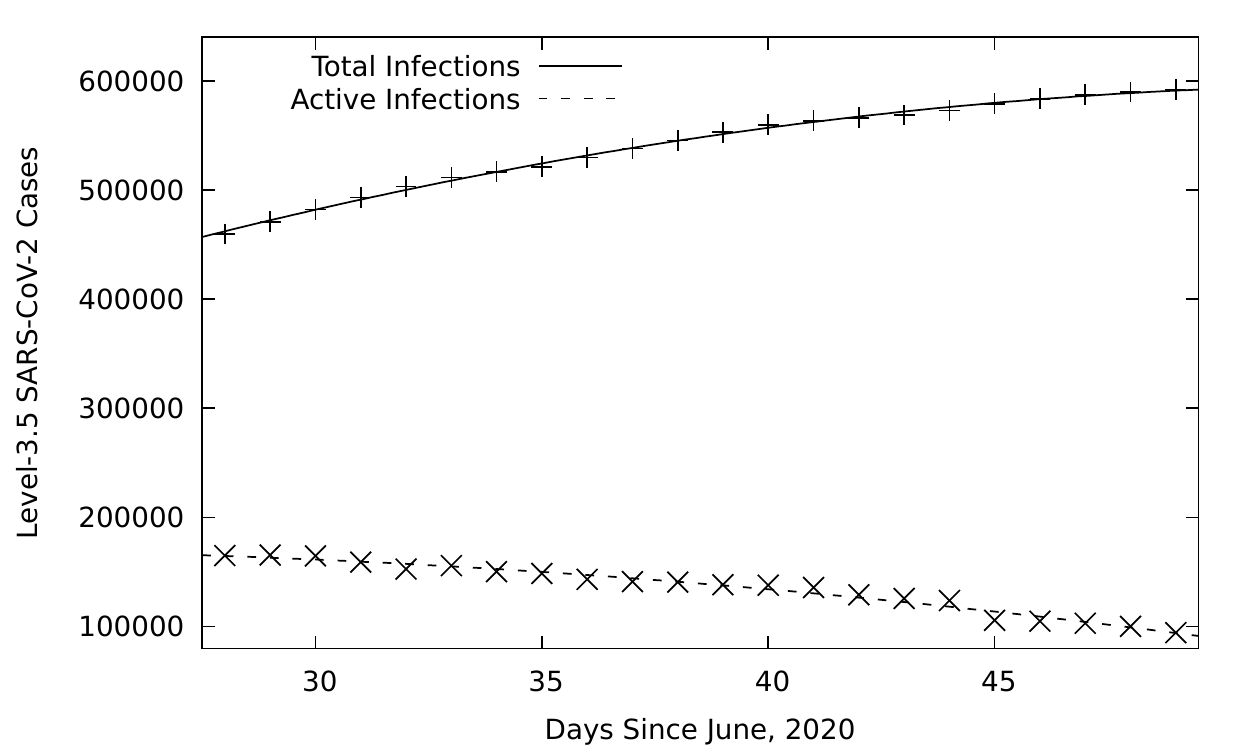}
	\caption{Level-3.5, best fits to \mbox{SARS-CoV-2}, infection data (28th of July to the 18th of August, 2020).} \label{level3.5}
   \end{center}
\end{figure}

The so-called level-3.5 lockdown commenced on the 13th of July and was in force up until the 17th of August. Once again, a period of 15 days was allowed for the viral incubation period and the subsequent diagnosis of an infection. Once again, it was also assumed that the termination of the level-3 lockdown would not reflect in the data for at least 24 hours. 
Curves were accordingly fitted to the subset of data (\cite{data}, \cite{openingWeeks} and \cite{data3}) which commenced on the 28th of July and terminated on the 18th of August. The curves fitted to the data, using Gnuplot, are depicted in Figure \ref{level3.5}.
The formula for the ``total infections'' curve is 
\[ 
%1.00372 \times 10^7 \ e^{0.000597062 \ t} - \ 9.73024 \times 10^6
-196.038 \ t^2 + 21230.6 \ t + 21520
\] 
and the formula for the ``active infections'' curve is 
\[ 
%5.2868 \times 10^6 \ e^{-0.000651511 \ t} - \ 5.02009 \times 10^6. 
-91.4826 \ t^2 + 3683.11 \ t + 133275
\] 
The values these formulae yield for the relevant dates are provided in \mbox{Table \ref{results}}. 

\subsection*{Converting the Conventional, Epidemiological Format into $S(t)$ and $I(t)$} 

The epidemiological data have been presented in their usual format ``numbers of
current infections'' and ``total number of cases''. If $N$ is the size of the
population, ``active infections'' are just ${\tilde I}(t) N$ and ``total
infections'' are just \mbox{$[{\tilde I}(t)+{\tilde R}(t) - {\tilde R}(0)] N$}.
Realising this, the values of $S(t)$ and $I(t)$ needed for this work can be
obtained from  
\begin{eqnarray*} \label{dataConversion1} 
S(t) &=& \frac{ N - a \times \mbox{total infections} } {N} - {\tilde R}(0) - b
\end{eqnarray*} 
and
\begin{eqnarray*} \label{dataConversion2} 
I(t) &=& \frac{a \times \mbox{active infections}}{N}. 
\end{eqnarray*} 
In 2020, the size of the South African population was estimated to be \mbox{59 140 502} by \cite{swedishPopulation}. 

\section{Calculation of the Basic Reproduction Number, $r_0$}

The basic reproduction number, $r_0$, was calculated according to the formula derived in \cite{childs} from Kermack and McKendrick's SIR equations \cite{kermackAndMcKendrick}. That is,
\begin{eqnarray} \label{formula}
r_0 = \frac{ {\mathop {\rm ln}} \left[ \frac{S(t_2)}{S(t_1)} \right] }{ \left[ I(t_2) + S(t_2) - I(t_1) - S(t_1) \right] },
\end{eqnarray}
in which $S(t)$, $I(t)$ and $R(t)$ represent the usual quantities in Kermack
and McKendrick's SIR model \cite{kermackAndMcKendrick} and they are evaluated at either end of an interval, $[t_1, t_2]$, in the formula. 

Expedience was the motivation for using this slightly unorthodox method,
however, not only have so-called individual level models, such as those of
\cite{andersonAndMay} and \cite{ilmVersusPlm}, been discredited by
\cite{ilmVersusPlm} as a means for calculating $r_0$, they are also far more
laborious than the simple method used in this work. 

Notice that this formula also appears to be fairly robust. Take any movement of
$I(t)$ by an additive constant, up or down, for example. Such movement has no
effect on the calculation of $r_0$, whatsoever. This is important as the
infection data, more than any other, are often plagued by exactly this problem.
In fact, the formula, Equation (\ref{formula}), is reasonably robust against any
data error that does not effect the relative values, or slopes of the functions
concerned. This will be borne out when exploring the use of a multiplicative
factor on the data.

\section{Calculation of the Threshold and $S_\infty$}

It is instructive to know both the threshold, as well as the point at which all
infection would cease. Both are calculated from $r_0$ and the relevant theory is
provided in \cite{childs}. If the susceptable fraction of the population is
still above $\frac{1}{r_0}$, for a given regime, the epidemic will continue to
grow in the way of infections. Only at the threshold does $r$-effective drop to
unity and the infections subsequently decline. Once $r_0$ has been calculated, $S(t_2) = S_\infty$, can be recovered from Equation (\ref{formula}), by considering that $S_\infty$ is the point at which all infection ceases, i.e. by setting $I(t_2)=0$, in Equation (\ref{formula}).

%Finally, the erroneously detected values of these selfsame quantities are predicted, should infections have been underestimated by a factor and an imperceptible, pre-existing, immune fraction of the population exist.

%In $r_0$ one therefore has a holistic quantity with which to characterise both the infectiousness of a disease, as well as the environment in which it propagates, right down to things like the temperature of nasal passages, the rules of a lockdown and even the level of non-compliance. The so-called $r$-effective, $r_0 \times S(t)$, is a characterisation of the disease's infectiousness, pertinent to a given point in time, as the epidemic progresses, or where immunity is present.

\begin{sidewaystable}
\begin{center}
\begin{tabular}{|c||c|c|c||c|c|c||c||c||c||} \hline & & & & & & & & & \\ Regime & Date 1 & Total Cases & Active Cases & Date 2 & Total Cases & Active Cases & $r_{0}$ & Threshold $S \ / \%$ & $S_\infty \ / \%$ \\ & & & & & & & & & \\ \hline
level 5 & 10.04 & 1 940 & 1 589 & 01.05 & 5 945 & 3 518 & 1.93 & 51.8 & 22.3 \\ \hline 
level 4 & 16.05 & 14 695 \ & 8 096 & 01.06 & 34 376 \ & 16 110 \ & 1.69 & 59.3 & 31.4 \\ \hline  
level 3 & 16.06 & 76 210 \ & 32 580 & 01.07 & 159 617 \ & 803 30 \ & 2.34 & 42.7 & 13.0 \\ \hline 
\texttt{"} & 16.06 & 75 493 \ & 30 996 & 13.07 & 289 922 \ & 146 472 \ & 2.17 & 46.0 & 16.2 \\ \hline 
level 3.5 & 28.07 & 462283 \ & 164 680 & 18.08 & 591 132 \ & 94 098 \ & 0.65 & --- & --- \\ \hline 
Sweden & 26.03 & 3 580 & 3 443 & 05.06 & 43 443 \ & 30 663 \ & 3.16 & 31.6 & \ 5.0 \\ \hline 
\end{tabular}
\caption{The scenario in which the data is assumed to be correct. The basic
reproduction number, $r_0$, the consequent threshold and the point at which all
infection ceases, $S_\infty$, have all been calculated with $a = 1$, $b=0$. With
the exception of Sweden, all are calculated from data associated with the
relevant lockdown regimes, imposed for the \mbox{SARS-CoV-2} pandemic, in South
Africa.} \label{results}
\vspace{20mm}
\begin{tabular}{|c||c|c|c|c||c|c|c||c|c||} 
\hline 
& & & & & & & & & \\ 
Regime & Date 1 & Date 2 & $a$ & $b$ & $r_{0}$ & Threshold $S \ / \%$ & $S_\infty \ / \%$ & Perceived Threshold ${\tilde S} \ / \%$ & ${\tilde S}_\infty \ / \%$ \\   
& & & & & & & & & \\ 
\hline
level 5 & 10.04 & 01.05 & 10 & 0 & 1.93 & 51.8 & 22.3 & 95.2 & 92.2 \\ \hline 
level 4 & 16.05 & 01.06 & \texttt{"} & \texttt{"} & 1.69 & 59.0 & 31.1 & {\bf 95.9} & 93.1 \\ \hline  
level 3 & 16.06 & 01.07 & \texttt{"} & \texttt{"} & 2.39 & 41.9 & 12.4 & 94.2 & 91.2 \\ \hline 
\texttt{"} & 16.06 & 13.07 & \texttt{"} & \texttt{"} & 2.24 & 44.7 & 15.0 & 94.5 & 91.5 \\ \hline 
level 3.5 & 28.07 & 18.08 & \texttt{"} & \texttt{"} & 0.71 & --- & --- & --- & --- \\ \hline   
\end{tabular}
\caption{The basic reproduction number, $r_0$, the consequent threshold, the
point at which all infection ceases, $S_\infty$, the erroneously detected
threshold and the erroneously detected point at which all infection ceases,
${\tilde S}_\infty$.}
\label{10results}
\end{center}
\end{sidewaystable}

\begin{sidewaystable}
\begin{center}
\begin{tabular}{|c||c|c|c|c||c|c|c||c|c||} 
\hline 
& & & & & & & & & \\ 
Regime & Date 1 & Date 2 & $a$ & $b$ & $r_{0}$ & Threshold $S \ / \%$ & $S_\infty \ / \%$ & Perceived Threshold ${\tilde S} \ / \%$ & ${\tilde S}_\infty \ / \%$ \\   
& & & & & & & & & \\ 
\hline
level 5 & 10.04 & 01.05 & 10 & 0.83 & 11.4 & 8.8 & 3.8 & 99.2 & 98.7 \\ \hline 
level 4 & 16.05 & 01.06 & \texttt{"} & \texttt{"} & 10.2 & 9.8 & 5.0 & {\bf 99.3} & 98.8 \\ \hline  
level 3 & 16.06 & 01.07 & \texttt{"} & \texttt{"} & 15.6 & 6.4 & 1.6 & 98.9 & 98.5 \\ \hline 
\texttt{"} & 16.06 & 13.07 & \texttt{"} & \texttt{"} & 15.7 & 6.4 & 1.6 & 98.9 & 98.5 \\ \hline 
level 3.5 & 28.07 & 18.08 & \texttt{"} & \texttt{"} & 8.03 & 12.5 & 5.4 & 99.5 & 98.8 \\ \hline   
\end{tabular}
\caption{The basic reproduction number, $r_0$, the consequent threshold, the
point at which all infection ceases, $S_\infty$, the erroneously detected
threshold and the erroneously detected point at which all infection ceases,
${\tilde S}_\infty$.}
\label{results2}
\vspace{20mm}
\begin{tabular}{|c||c|c|c|c||c|c|c||c|c||} 
\hline 
& & & & & & & & & \\ 
Regime & Date 1 & Date 2 & $a$ & $b$ & $r_{0}$ & Threshold $S \ / \%$ & $S_\infty \ / \%$ & Perceived Threshold ${\tilde S} \ / \%$ & ${\tilde S}_\infty \ / \%$ \\ & & & & & & & & & \\ 
\hline
level 5 & 10.04 & 01.05 & 20 & 0.65 & 5.53 & 18.1 & 7.7 & 99.2 & 98.6 \\ \hline 
level 4 & 16.05 & 01.06 & \texttt{"} & \texttt{"} & 4.94 & 20.3 & 10.3 & {\bf 99.3} & 98.8 \\ \hline 
level 3 & 16.06 & 01.07 & \texttt{"} & \texttt{"} & 7.55 & 13.2 & 3.3 & 98.9 & 98.4 \\ \hline 
\texttt{"} & 16.06 & 13.07 & \texttt{"} & \texttt{"} & 7.56 & 13.2 & 3.3 & 98.9 & 98.4 \\ \hline 
level 3.5 & 28.07 & 18.08 & \texttt{"} & \texttt{"} & 3.78 & 26.5 & 11.8 & 99.6 & 98.8 \\ \hline   
\end{tabular}
\caption{The basic reproduction number, $r_0$, the consequent threshold, the
point at which all infection ceases, $S_\infty$, the erroneously detected
threshold and the erroneously detected point at which all infection ceases,
${\tilde S}_\infty$.}
\label{results3}
\end{center}
\end{sidewaystable}

\begin{sidewaystable}
\begin{center}

\begin{tabular}{|c||c|c|c|c||c|c|c||c|c||} 
\hline 
& & & & & & & & & \\ 
Regime & Date 1 & Date 2 & $a$ & $b$ & $r_{0}$ & Threshold $S \ / \%$ & $S_\infty \ / \%$ & Perceived Threshold ${\tilde S} \ / \%$ & ${\tilde S}_\infty \ / \%$ \\   
& & & & & & & & & \\ 
\hline
level 5 & 10.04 & 01.05 & 30 & 0.5 & 3.87 & 25.8 & 11.0 & 99.2 & 98.7 \\ \hline 
level 4 & 16.05 & 01.06 & \texttt{"} & \texttt{"} & 3.46 & 28.9 & 14.7 & {\bf 99.3} & 98.8 \\ 
\hline  
level 3 & 16.06 & 01.07 & \texttt{"} & \texttt{"} & 5.32 & 18.8 & 4.7 & 99.0 & 98.4 \\ 
\hline
\texttt{"} & 16.06 & 13.07 & \texttt{"} & \texttt{"}  & 5.35 & 18.7 & 4.6 & 99.0 & 98.5 \\ 
\hline 
level 3.5 & 28.07 & 18.08 & \texttt{"} & \texttt{"} & 2.79 & 35.8 & 15.4 & 99.5 & 98.8 \\ 
\hline 
\end{tabular}
\caption{The basic reproduction number, $r_0$, the consequent threshold, the
point at which all infection ceases, $S_\infty$, the erroneously detected
threshold and the erroneously detected point at which all infection ceases,
${\tilde S}_\infty$.} \label{results4}
\vspace{20mm}
\begin{tabular}{|c||c|c|c|c||c|c|c||c|c||} \hline & & & & & & & & & \\ Regime & Date 1 & Date 2 & $a$ & $b$ & $r_{0}$ & Threshold $S \ / \%$ & $S_\infty \ / \%$ & Perceived Threshold ${\tilde S} \ / \%$ & ${\tilde S}_\infty \ / \%$ \\    
& & & & & & & & & \\ 
\hline
level 5 & 10.04 & 01.05 & 60 & 0 \ & 1.94 & 51.6 & 22.1 & 99.2 & 98.7 \\ 
\hline 
level 4 & 16.05 & 01.06 & \texttt{"} & \texttt{"} & 1.73 & 57.8 & 29.4 & {\bf 99.3} & 98.8 \\ \hline  
level 3 & 16.06 & 01.07 & \texttt{"} & \texttt{"} & 2.66 & 37.6 & 9.3 & 99.0 & 98.5 \\ \hline
\texttt{"} & 16.06 & 13.07 & \texttt{"} & \texttt{"} & 2.68 & 37.4 & 9.2 & 99.0 & 98.5 \\ \hline 
level 3.5 & 28.07 & 18.08 & \texttt{"} & \texttt{"} & 1.40 & 71.6 & 30.8 & 99.5 & 98.8 \\ \hline 
\end{tabular}
\caption{The basic reproduction number, $r_0$, the consequent threshold, the
point at which all infection ceases, $S_\infty$, the erroneously detected
threshold and the erroneously detected point at which all infection ceases,
${\tilde S}_\infty$.} \label{results5}
\end{center}
\end{sidewaystable}

\section{Results}

Since the rules in place at the time of the peak were most similar to level 4,
the values of $b$ and $a$ explored were mostly selected on the basis that they
manufacture an erroneously-detected, \mbox{99.3 \%} threshold for level 4. The
results, as well as the inputs from which they were obtained, are provided on
pages \pageref{results} \mbox{to \pageref{results5}}. 

\section{Conclusions}

The \mbox{SARS-CoV-2} pandemic was, very likely, a lot closer to its threshold
than the South African data suggested, however, some, possibly external, factor
still changed the value of $r_0$. The author's opinion is therefore that the
contemplated data-deficiencies are unlikely to explain the early peak in the
\mbox{SARS-CoV-2} pandemic on their own. If deficient data did, indeed, play a
role, then the more compelling of the two phenomena might be that the true level
of infection has been underestimated by a multiplicative factor. It is already a
documented phenomenon and it is mathematically less disruptive. The existance of
a significant, imperceptible, immune fraction of the population quickly drives
$r_0$ up, admittedly not necessarily to unprecedented levels, however, in so
doing, it moves thresholds to even lower values. In contrast, there is very
little difference between perceived and true $r_0$'s, should infections have
been underestimated by a factor. It is also already a widely recognised
phenomenon that the \mbox{SARS-CoV-2} virus is often insidious and infection
data are incorrect by a substantial factor, whereas a pre-existing,
imperceptible immune fraction of the population remains nothing more than a
hypothetical construct; a mere thought experiment for the present. 

The phenomenon of infections having been underestimated by a multiplicative
factor, alone, is unable to comprehensively explain the \mbox{SARS-CoV-2} peak
observed in the South African data, without contemplating improbably-high
values. Yet, those improbably-high values ($a = 60$) might be possible in an
area like Khayalitsha, considering that a massive 57 \% of the inhabitants of
Mumbai slum areas tested positive \cite{mumbai}. Revising country-wide
infections upward by a single order of magnitude is probably something not too
far-fetched and it creates a level-4 threshold of 59.0 \% that would be
erroneously-detected as being 95.9 \%, a perceived value not too far from the
actual 99.3 \% peak observed (see Table \ref{10results}). It comes a long way to
reconciling the observed with the predicted and it remains an inescapable fact
that, in July, $r_0$ did change fairly abruptly in all the contemplated
scenarios of data-deficiency. This is not something suggestive of a threshold.

Around the 13th of July, it seems likely that something changed the value of
$r_0$ significantly. Perhaps one criticism of the \cite{childs} analysis of
lockdown regimes is that it denies the public at large their humanity. The
quantification of lockdown regimes means nothing if the rules aren't complied
with. Perhaps, by the 13th of July, the population's assymptotic compliance had
reached the necessary level? Perhaps, too, the intra-household infections that
obviously took place early on in the lockdowns, had also run their course? How
much genetic drift took place in the preceding months is yet another unanswered
question. Perhaps, when the prohibition was reimposed, the seemingly-endless
supply of liquor, that had understandably leaked from the stricken hospitality
industry, had finally dried up? All these factors could have contributed to
lowering the level-3.5 $r_0$ to below the level-4 value of 1.69, driving the
erroneously-detected 95.9 \% (Table \ref{10results}) closer to the 99.3 \% peak
observed. However, could all of this have driven the 1.69 down as far as 1.08,
the $r_0$-value necessary for the 99.3 \% perception? The fact remains that
level 3's $r_0$ was very much lower than 1.08. It was 0.71 for $a = 10$ and
$b=0$ (Table \ref{10results})! This was no threshold.

One external factor possibly worth considering is the change in weather that
coincided with the reinstatement of the prohibition and curfew. On the 13th of
July, winter arrived in force, imposing a lockdown of its very own. A massive
frontal system made its landfall, bringing snow to six provinces of what is,
traditionally, a warm country. Braais were called off. Many shack-dwellers were
confined to their blankets. Large groups no longer gathered on street corners
and outside certain houses to `bounce' liquor and cigrettes. The streets emptied
out and finally became deserted. No late-night figure even darted between the
throbbing blue houses. People wore their masks to stay warm. Exactly five days
later, on the 18th of July, the incidence of \mbox{SARS-CoV-2} cases peaked (the
mean \mbox{SARS-CoV-2} incubation period is 5.2 days
\cite{meanIncubationPeriod}) and the number of active infections followed suit
on the 23rd of July. Could something so simple be the explanation? The author
is, otherwise, at a loss. It should be borne in mind that the model used is
ideally intended for homogenous populations, settlements like Khayalitsha, not
really whole countries.

\bibliography{covidLockdownSequalBibliography}

\end{document}